\documentclass[prd,twocolumn]{revtex4}
\usepackage{epsfig}
\usepackage{amsmath}
\newcommand{\be}{\begin{equation}}
\newcommand{\ee}{\end{equation}}
\newcommand{\bm}{\begin{mathletters}}
\newcommand{\eem}{\end{mathletters}}
\newcommand{\bes}{\begin{subequations}}
\newcommand{\ees}{\end{subequations}}

\newcommand{\bi}{\begin{itemize}}
\newcommand{\ei}{\end{itemize}}

\begin{document}
\title{A New Strong Interaction Sector as the origin
for the Dark Energy and Dark Matter}
\author{P.Q. Hung}
\email[]{pqh@virginia.edu}
\affiliation{Dept. of Physics, University of Virginia, \\
382 McCormick Road, P. O. Box 400714, Charlottesville, Virginia 22904-4714, 
USA}
\date{\today}
\begin{abstract}
A new strong interaction sector is proposed as a possible origin
for the Dark Energy and Dark matter. It is described by an unbroken
gauge group $SU(2)_Z$ which grows strong at a scale $\Lambda_Z
\sim 10^{-3}\,eV$ provided its coupling is of the order of the
electroweak coupling at some high energy scale (either
$\sim 10^{16}\,GeV$ or $\sim O(TeV)$ . The present vacuum 
energy which is consistent with
a cosmological constant is related to $\Lambda_Z$ and one
of the $SU(2)_Z$ ``matter fields'' provides a candidate for the Dark Matter
in the form of a weakly interacting massive particle (WIMP)
with an annihilation cross section naturally of the order of the
electroweak one at the time of its decoupling.
\end{abstract}
\maketitle

The nature of the dark matter and dark energy 
(responsible for an accelerating
universe \cite{acceleration}) is one of the deepest problem in 
contemporary cosmology \cite{FT}.
With the equation of state relating the pressure to the
energy density being $p=w\rho$, the latest CTIO Lensing 
Survey \cite{CTIO} performed in conjunction
with CMB and Type Ia supernovae data up to redshift $z \sim 0.4$
found a value of $w$ which is consistent with $-1$. This,
in turn, is consistent with the scenario in which the present
total energy density is dominated by the vacuum energy or
cosmological constant.
However, it is well-known that this vacuum energy is 
surprisingly tiny, namely $\rho_{V} 
\approx (10^{-3}\,eV)^4$. We shall take the point of view that
this could represent a {\em new scale of physics} on the
same footing as other known scales ($\Lambda_{EW} \sim
250\,GeV$, $\Lambda_{QCD} \sim 200\,MeV$) rather than 
one in which the associated vacuum energies 
{\em so finely cancel} down to $(10^{-3}\,eV)^4$. (We assume
that the electroweak and QCD vacuum energies are set to zero
after the associated phase transitions are completed.)
The purpose
of this paper is to propose a new interaction which grows
{\em strong} at $\Lambda_{Z} \sim 10^{-3}\,eV$ with interesting
implications for the Dark Energy and Dark Matter.

The model we would like to propose is based on a vector-like 
unbroken $SU(2)_Z$ gauge group whose coupling starts out
with
a value assumed to be close to that of the SM $SU(2)_L$
at some high energy scale and 
becomes large at $\sim 10^{-3}\,eV$. 
Here the subscript $Z$ refers to an
ancient greek word {\em zophos} which means {\em darkness}.
Hence, this new interaction will be referred to
as Quantum Zophodynamics or QZD.
The  $SU(2)_Z$ matter fields are Standard Model (SM) singlets
and vice versa.
As such, the two sectors would be {\em incommunicado}
(except through gravity) if not for the fact they 
can communicate 
with each other, albeit {\em very weakly}, through
a ``messenger'' field which carries both SM and
$SU(2)_Z$ quantum numbers. 


Our model was based on a very simple observation: For an unbroken gauge group
(with a coupling comparable in magnitude to the $SU(2)_L$ gauge
coupling at high energy) to become confining at energy scales many orders
of magnitude below the QCD scale, its 
one-loop beta function has to be small compared with that of
$SU(3)_c$. 

We propose:

\bi

\item An unbroken vector-like gauge group $SU(2)_Z$.

\item The initial value of the $SU(2)_Z$ gauge coupling is
similar to that of the SM $SU(2)_L$ coupling at some high energy scale
such as the early unification scale (PUT)\cite{BH} in the TeV range
(or even the GUT scale of O($10^{16}$ GeV). This assumption
makes the model fairly {\em predictive} as we shall see below
(although a more general case can be easily entertained and
will be dealt with elsewhere).

\item Fermions (or chiral superfields in the SUSY case)
transform as adjoints of $SU(2)_Z$ and
singlets under the SM. We denote them by
$\psi^{(Z)}_{L,R} = (1,1,0,3)$ under $ SU(3) \otimes SU(2)_L
\otimes U(1)_Y \otimes SU(2)_Z$ for the
non-SUSY case, and similarly $\Phi^{(Z)}_{L,R} = (1,1,0,3)$ for the
chiral superfield. These are all {\em electrically neutral}
and the fermions would be qualified for the designation ``sterile
neutrinos''.

\item Messenger fields transform either as fundamentals
{\em or} as adjoints under $SU(2)_Z$ 
and possess SM quantum numbers. They are denoted by
$\varphi_{1/2}^{(Z)} =(\varphi_{1/2}^{(Z),0},\varphi_{1/2}^{(Z),-})=
(1,2,Y_{\varphi}=-1,\,2)$ or 
$\varphi_{1}^{(Z)} =(\varphi_{1/2}^{(Z),0},\varphi_{1/2}^{(Z),-})=
(1,2,Y_{\varphi}=-1,\,3)$ 
under $SU(3) \otimes SU(2)_L
\otimes U(1)_Y \otimes SU(2)_Z$ for the non-SUSY case and
$\chi_{L,R} =(1,1,Y_{\chi},2)$ for the SUSY case (the triplet
case is not allowed here if we want the model to be
asymptotically free).


\ei

We split the discussion into two parts: Particle Physics and Cosmology.


{\bf (A) PARTICLE PHYSICS}:

The basic one-loop RG equation is $\frac{\mu\,d\alpha_Z}{d\mu} = 
-\frac{\beta_0}{2\,\pi} \alpha_{Z}^2$
where $\beta_{0,NS} = (11C_{2}(G)/3 -2/3\,\sum T_{i}(R)
-1/3\,\sum T_{S}(R))$ for the non-SUSY case and
$\beta_{0,S} = (3C_{2}(G) -\sum T_{\chi}(R))$ for the SUSY case,
with $T_{i}(R), T_{S}(R),T_{\chi}(R))$ for chiral fermion,
scalar, and chiral superfield respectively. 
From the particle content listed above,
one has $C_{2}(G)=T_{\psi^Z}=T_{\Phi}=2$, $T_{\varphi}=
T_{\chi}=1/2$ or $T_{\varphi}=2$ (for the
Georgi-Glashow type of situation), giving
$\beta_{0,NS} =  (22-8 n_{\psi} -n_{\varphi})/3$
or $\beta_{0,NS} =  (22-8 n_{\psi} -4n_{\varphi})/3$, 
for the non-SUSY case, and
$\beta_{0,S} =  (6-4\,n_{\Phi}-n_{\chi})$, 
for the SUSY case. Here, $n_{\psi}$, $n_{\Phi}$,
and $n_{\chi}$ already include {\em both left and right-handed fields}.
We wish to make $\beta_0$ as small as
possible while keeping it positive. This leads to the following
minimal set:
(I) $n_{\psi}=2\,;\, n_{\varphi}=1$,
for the non-SUSY case and
(II) $n_{\Phi}=1\,;\, n_{\chi}=1$,
for the SUSY case. Above
all mass thresholds, one has $\beta_{0,NS}=5/3$ or $\beta_{0,NS}=2/3$, 
$\beta_{0,S} =1$,
to be compared with $\beta_{0,SU(2)_L}=10/3$ and $\beta_{0,SU(3)_c} =7$
in the SM with three families. 

Below, we will present a few numerical results, leaving a general
analysis for a longer version of the paper.

{\bf (I) Non-SUSY $SU(2)_Z$}:

Since $\varphi$ also carries the SM quantum number, it is safe and not
unreasonable to assume it has a mass of the order of the electroweak
scale $\Lambda_{EW} \approx 250 GeV$. 
We will assume
that the potential for $\varphi$ is such that $<\varphi>=0$
in order for $SU(2)_Z$ to be {\em unbroken}.

For the fermions $\psi^{(Z)}_{L,R}$ which are singlets with respect
to the SM model, one can write down a gauge-invariant mass term of the
type $\sum_{i=1}^{i=2} m_{i} \bar{\psi}^{(Z)}_{i}\,\psi^{(Z)}_{i}$.
For the purpose of this paper, we will deal with the case 
where $m_1 \ll m_2$.


Let $M$ be the starting scale (early unification scale or GUT).
The four basic equations which we will use are the following:
1) $\alpha_Z(\Lambda_{EW}) = \alpha_Z(M)/\{1+(\alpha_Z(M)/2\pi)
(K_2\,or K_3)\ln(\Lambda_{EW}/M)\}$ where $K_2 =5/3$
for $\varphi_{1/2}^{(Z)}$
and $K_2 =2/3$ for $\varphi_{1}^{(Z)}$;
2)
$\alpha_Z(m_2) = \alpha_Z(\Lambda_{EW})/\{1+(\alpha_Z(\Lambda_{EW})/\pi)
\ln(m_2/\Lambda_{EW})\}$; 3)
$\alpha_Z(m_1) = \alpha_Z(m_2)/\{1+(7 \alpha_Z(m_2)/3\,\pi)
\ln(m_1/m_2)\}$; 4)
$\Lambda_{Z}=m_{1} \exp\{-(3\pi/11\alpha_Z(m_1))(1-\alpha_Z(m_1))\}$.
If we assume $m_1=0$, then one can skip (3) and replace (4) by
$\Lambda_{Z}=m_{2} \exp\{-(3\pi/7\alpha_Z(m_2))(1-\alpha_Z(m_2))\}$.
 
For comparison, let us recall that the $SU(2)_L$
coupling at $M_Z$ is $\alpha_{2}^{-1}(M_Z) \sim 29.6$. At
$2\,TeV$, it is roughly (depending on unknown particle threshold)
$\alpha_{2}^{-1}(2\,TeV) \sim 28.2$. At $M= 2 \times 10^{16}\,GeV$,
$\alpha_{2}^{-1}(M) \sim 45$ for the non-SUSY case and
$\alpha_{2}^{-1}(M) \sim 24$ for the SUSY case with three
generations.

In Table I, we list the results coming from the requirement:
$\alpha_{Z}(\Lambda_Z) \sim 1$ at $\Lambda_{Z} \sim 10^{-3}\,eV$,
for doublet (bi-fundamental) $\varphi_{1/2}^{(Z)}$ and 
triplet $\varphi_{1}^{(Z)}$.

\begin{table}
\caption{\label{tab:table1} Correlations between $m_1$, $m_2$
and $\alpha_{Z}^{-1}(M)$} 
\begin{ruledtabular}
\begin{tabular}{ccccc} 
&$m_1$&$m_2$&$\alpha_{Z}^{-1}(M)$ \\ \hline
GUT ($\varphi_{1/2}^{(Z)}$): $M= 2 \times 10^{16}\,GeV$&
$0$&$30\,MeV$&$30$ \\
&$1.5\,eV$&$30\,MeV$&$33$ \\
&$0$&$200\,GeV$&$34$ \\
&$1\,eV$&$200\,GeV$&$36.5$ \\ \hline
GUT ($\varphi_{1}^{(Z)}$): $M= 2 \times 10^{16}\,GeV$&
$1\,eV$&$200\,GeV$&$31.5$ \\ \hline
PUT ($\varphi_{1/2}^{(Z)}$): $M= 2\,TeV$&
$0$&$30\,GeV$&$25$ \\
&$1\,eV$&$30\,GeV$&$27.8$ \\
&$0$&$120\,GeV$&$25.6$ \\
&$1\,eV$&$120\,GeV$&$28.4$ \\ \hline
PUT ($\varphi_{1}^{(Z)}$): $M= 2\,TeV$&
$0$&$200\,GeV$&$25.6$ \\
&$1\,eV$&$200\,GeV$&$28.2$ \\
\end{tabular}
\end{ruledtabular}
\end{table}


From Table I, one notices
how important it is to have a ``large'' value for
$m_2$, a non-zero value for $m_1$ and to have the presence of 
$\varphi^{(Z)}$ 
if we wish $\alpha_Z(M)$ to have a value close to the SM one.

{\bf (II) SUSY $SU(2)_Z$}:

We will assume the mass of the $\chi$ superfield
is of the order of the electroweak scale. 
We will further
assume that SUSY breaking is such that the scalar field
inside $\Phi^{(Z)}_{L,R}$ has a mass $m_S$ while the
fermionic partner remains massless. 
We obtain:
\{$\alpha^{-1}_Z(M)=22.5$; $m_S=1.5\,GeV$\},
\{$\alpha^{-1}_Z(M)=1/23.5$; $m_S=150\,GeV$\},
for $M= 2 \times 10^{16}\,GeV$, $\Lambda_{Z} \sim 1.4 \times
10^{-3}\,eV$,
to be compared with the typical SUSY GUT 
scale $\alpha^{-1}_{GUT} \sim 24$.

{\bf (III) Comments}:

Although the above results are but a sample of a more comprehensive
analysis, a pattern clearly emerges: The $SU(2)_Z$ gauge coupling
can start out at high energy with a magnitude {\em close} to 
the value of the $SU(2)_L$ coupling and subsequently
becomes large at a scale
which is {\em eleven orders of magnitude smaller than the QCD scale}.
When the lighter of the two $\psi^{(Z)}$ has a mass at around
$1\,eV$ and the heavier one with a mass of O($100\,GeV$), the
initial $SU(2)_Z$ gauge coupling is {\em remarkably} close
in value to that of $SU(2)_L$. If one simply starts
out with the assumption $\alpha_Z(M) \sim \alpha_2(M)$,  
the sought-after scale
$\Lambda_{Z} \sim  10^{-3}\,eV$ arises {\em naturally}. 
Notice that
$\alpha_Z(M)$ is much closer to $\alpha_2(M)$ for the
early unification scenario \cite{BH} than for GUT as can be seen
from Table I. Furthermore, this
scenario is closer in spirit and in construction to the early unification
picture \cite{BH} where the gauge couplings of the various $SU(2)$'s
are assumed to be equal at the PUT unification scale of the order of
a few TeVs. 

{\bf(IV)  Messenger fields and their implications}:

The manner in which the $SU(2)_Z$ matter communicates with normal
matter will depend on whether $\varphi^{(Z)}$ is a doublet or
a triplet of $SU(2)_Z$ for the non-SUSY scenario. 

For $\varphi_{1}^{(Z)}$, one can write down a gauge-invariant
Yukawa coupling between $\varphi_{1}^{(Z)}$, the
normal leptons, and $\psi_{2}^{(Z)}$ (and {\em not}
$\psi_{1}^{(Z)}$, the reason of which will become clear below) as:
\be
\label{yukvarphi}
{\cal L}_{\varphi_{1}} = \sum_{i} g_{Yi}\,\bar{l}_{L}^{i}\,
\varphi_{1}^{(Z)}\,\psi^{(Z)}_{2,R}+ h.c. \,,
\ee
where $l_L$ is a normal lepton doublet and the sum is over
three generations. (This could be accomplished by assuming
a discrete symmetry in such a way that a similar coupling to
$\psi_{1}^{(Z)}$ is forbidden.)
A scattering process of the type
$l_i+\psi^{(Z)} \rightarrow \psi^{(Z)} +l_j$ via a heavy
$\varphi_{1}$ has an amplitude proportional to
$g^{*}_{Yi}g_{Yj}/m_{\varphi^{(Z)}}^2$. What the constraints
on the arbitrary Yukawa couplings $g_{Yi}$ might be is of
great interest for the Dark Matter search.

For $\varphi_{1/2}^{(Z)}$ (as well as for
the SUSY $\chi_{L,R}=(1,1,Y_{\chi},2)$), no such Yukawa coupling
can be written down and the interaction of $SU(2)_Z$ matter
with normal matter proceeds through
its anomalous magnetic moment (obtained at two loops)
$\mu^{(Z)}_{1,2} \sim [2\,m_{e}\,m_{1,2}/m_{\varphi^{(Z)}}^2]
[\alpha_{Z}^2 \,/16 \pi^2]\times log terms \times \mu_{B}$,
where $\mu_{B}$ is the Bohr magneton. (Recall that, in this case,
$SU(2)_Z$ matter interacts with the messenger field by exchanging
$SU(2)_Z$ ``gluons''.)
Since $m_1 \sim 1\,eV$, 
one can see
that $\mu^{(Z)}_{1}$ is completely negligible. However,
$\mu^{(Z)}_{2}$ can be as large as $10^{-8}\,\mu_B$ (in the
regime where $\alpha_{Z} =1$).
The Lagrangian for the interaction between
$\psi_{1,2}^{(Z)}$ and the photon can be written as \cite{krasnikov}
${\cal L}_{\psi_{1,2}^{(Z)}} = \mu^{(Z)}_{1,2}\,\bar{\psi_{1,2}^{(Z)}}
(\sigma_{\mu\nu}/2)\,\psi_{1,2}^{(Z)}\,F^{\mu\nu} + h.c.$
Notice that $\varphi_{2}^{(Z)}$ cannot decay into lighter normal
SM particles, neither through Eq. (\ref{yukvarphi}) nor through
its anomalous magnetic moment. As a result it is absolutely
{\em stable}

The messenger field $\varphi^{(Z)}$ carrying {\em both} $SU(2)_Z$
and SM quantum numbers can be searched for at colliders such as
the LHC.
As shown below,
$\psi_{2}^{(Z)}$ has all the properties for being a Dark Matter
candidate in the form of a WIMP. The previous discussion will
be useful in the search for the Dark Matter, the
phenomelogy of which will be presented elsewhere. Notice also
that, since both $\psi_{1,2}^{(Z)}$ are ``sterile neutrinos'',
$\psi_{1}^{(Z)}$ with a mass around $1\,eV$ looks fairly
suggestive of the LSND \cite{LSND} one, although it remains
to be seen if this is the case. 


{\bf (B) COSMOLOGICAL IMPLICATIONS}:


{\bf (I) $SU(2)_Z$ ``brief thermal history''}:

At $T \gg m_i$, where $m_i$ is a generic particle mass, all
normal matter and matter that carries $SU(2)_Z$ quantum numbers
are in thermal equilibrium and are characterized by
a common temperature. The fact that $SU(2)_Z$ matter is
in thermal equilibrium with normal matter is because the
messenger fields, e.g. $\varphi^{(Z)}$, carry both SM and
$\varphi^{(Z)}$ quantum numbers and, therefore, can interact
with normal matter as well as with the $SU(2)_Z$ ``gluons'' and
fermions $\psi^{(Z)}$. (Let us recall
that $\varphi^{(Z)}$ also serves
the purpose of slowing the evolution of the $SU(2)_Z$ coupling
so that it can grow strong at $\sim meV$ if 
$\alpha_Z(M) \sim \alpha_2(M)$.) 

When $T<m_{\varphi^{(Z)}}$, the 
now-non-relativistic messenger field $\varphi^{(Z)}$ begins
to annihilate each other with their number density
decreasing like $n \sim (m_{\varphi^{(Z)}}\,T)^{3/2}
\exp(-m/T)$. For the charged messenger
$\varphi^{(Z),+}$, the dominant annihilation cross section
is the electromagnetic one which goes like $\sigma \sim
\alpha^2/T^2$ which grows with decreasing $T$. Their relic
density would be negligible. For $\varphi^{(Z),0}$, it can
decay into either 2 $Z$ bosons or into lighter SM particles
via 2 $Z$ bosons. Again, its present relic density would be 
negligible.

For $\psi_{2}^{(Z)}$ with mass $m_2$, the situation becomes
interesting when $T<m_2$. Notice that, since 
$m_{\varphi^{(Z)}} > m_2$, $\psi_{2}^{(Z)}$ is a 
{\em stable particle}. 
The dominant annihilation cross section
(into 2 $\psi_{1}^{(Z)}$ or 2 $SU(2)_Z$ gluons) 
behaves as $\sigma_{Z} \sim \alpha_{Z}^{2}(T)/m_{2}^2$. 
As we have seen above, the
crucial feature of the $SU(2)_Z$ model is that its coupling
basically almost ``parallels'' that of the SM $SU(2)_L$.
From the above analysis, one can make the
approximation $\alpha_Z(T \sim M_Z) \sim \alpha_{2} (T \sim M_Z)$.
This implies that $\sigma_{Z}$ is naturally of the order of the
weak cross section for $m_2 = O(100\,GeV)$. This will have
an interesting implication concerning the nature of the dark
matter as shown below. 

After $\varphi^{(Z)}$ and $\psi_{2}^{(Z)}$ decouple, the only
relativistic particles left in the $SU(2)_Z$ sector are
the $SU(2)_Z$ 'gluons'' and $\psi_{1}^{(Z)}$. As we have
mentioned earlier, these particles interact {\em very weakly} with
normal matter and {\em decouple soon} after $\varphi^{(Z)}$ and 
$\psi_{2}^{(Z)}$ went out of equilibrium. Their temperature
$T_Z$ would go like $R^{-1}$. It is important to compare
$T_Z$ with $T$ for the normal matter at the time of Big Bang
Nucleosynthesis (BBN) \cite{kolb}. The SM number of degrees of freedom
(including $\varphi^{(Z)}$) for three families above the top quark
mass is $g_{\ast} = 427/4 + 8= 459/4$ and between $m_e$ and 
$m_{\mu}$ is $g_{\ast}^{'} = 43/4$. Because of entropy
conservation, the relationship between $T_Z$ and $T$
at $m_e < T < m_{\mu}$ is simply $T_Z = (43/459)^{1/3}T
\sim 0.45\,T$. During that period, $g_{\ast}^{SU(2)_Z}=
33/2$ and $\rho_{SU(2)_Z}/\rho_{SM} = (66/43)(43/459)^{4/3}
\sim 0.07$ and one can see that the presence of $SU(2)_Z$
``matter'' will have a negligible effect on BBN. After $e^{\pm}$
decoupling, one obtains $T_Z = [(43/459)(4/11)]^{1/3}T
\sim T/3$. 

{\bf (II) Dark Energy}:

We now discuss two possible scenarios: one having to do with
a first-order phase transition from an unbroken 
chiral symmetry to a broken
one, and another with an axion-like potential for $SU(2)_Z$.
If the present universe is dominated by a vacuum energy of
the order of $(\Lambda_Z)^4 \sim (10^{-3}\,eV)^4$ and is
entering an inflationary stage, how long will it last? The first-order
phase transition proceeds by the nucleation of bubbles of the
true vacuum with a rate $\Gamma = A \exp\{-S_{E}\}$, where $A$
is a prefactor and $S_E$ is the Euclidean action. If this rate
is small, the inflationary period can last for a long time.
(Notice also that $SU(2)_Z$ provides a rationale for the scale
in front of the flat potential of \cite{hung}.) 

{\bf (a) Chiral phase transition}:

Extensive studies of
chiral phase transition in QCD have been performed and they
appeared to suggest that it is of first-order. One might
expect a similar first-order phase transition for
$SU(2)_Z$. 
There have also been studies of the
chiral phase transition by bubble nucleation, e.g. one in which
the linear $\sigma$ model is used \cite{chi}, and there it is
seen that, at $T=T_c$, the bubble nucleation
rate is negligible. A similar study for the $SU(2)_Z$ case is
under investigation. 
Since one scenario for 
the messenger field involves $\varphi^{(Z)}$ in the adjoint 
representation, one can also investigate
the phase structure of the Georgi-
Glashow (Polyakov) model in 2+1 dimensions \cite{antonov}.

{\bf (b) Axion-like potential}:

There has been interesting proposals to use the QCD (or QCD-like) axion 
\cite{axion} potential $V(a) =V_{0}[1-\cos\frac{Na}{v}] -
\eta\cos[\frac{a}{v} +\gamma]$ ($V_0 \sim \Lambda_{QCD}^4$)
as a model for the early inflationary scenario \cite{freese}.
We would like to propose a similar scenario but, this time, as an
explanation for the origin of the dark energy. The $SU(2)_Z$
axion comes from a $U(1)_{``PQ''}$-like symmetry of the $SU(2)_Z$
fermions. For the purpose of this paper, we will only need
to consider the case of the $SU(2)_Z$ ({\em zophos}) instanton
potential with $N=1$. The $SU(2)_Z$ axion $a_Z$ (where 
$\phi_Z = \frac{1}{\sqrt{2}}(v + \rho) \exp(i\frac{a_Z}{v})$) has
a periodicity $a_Z \rightarrow a_Z + 2\pi\,v = a_Z + 2 \pi f_a$, 
where we assume
that the axion decay constant $f_a = O(v)$. (For this analysis, it
is irrelevant whether or not $f_a=v$ as long as they are of the same 
order.) We write
\be
\label{potential}
V(a_Z) = \Lambda_Z^4 [1-\cos\frac{a_Z}{v}] -\epsilon\,\frac{a_Z}{2\pi v}\,.
\ee
In Eq.(\ref{potential}), the first term on the right hand side
represents the potential with two degenerate vacuua and the
second term represents a soft-breaking term lifting that degeneracy.
The difference in energy density between the two vacuua is $\epsilon$.
We will assume that $\epsilon \leq \Lambda_Z^4$ but not too different
from it and at the same time have a small barrier between the two vacuua. 
In other words we assume $\epsilon = O(\Lambda_Z^4)$ and
try to see its consequence on the cosmology of the model. The false
vacuum of the ``axion'' potential is at $a_Z = 2 \pi v$ and the
true vacuum is at $a_Z=0$. The scenario goes as follows.

(i) As $T_Z$ drops below $T_c \sim \Lambda_Z \sim 10^{-3}\,eV$, the universe
is trapped in the false vacuum with $a_Z = 2 \pi v$. The total energy
density is dominated by $\epsilon = O(\Lambda_Z^4)$. 

(ii) The first order phase transition to the true vacuum at $a_Z=0$
proceeds by bubble nucleation. 
The Euclidean action $S_E$, in the thin wall limit, can be
computed by looking at $\tilde{S}=\int_{a_Z = 2 \pi v}^{a_Z=0}
\sqrt{2\,\Lambda_Z^4\,[1-\cos\frac{a_Z}{v}]} da_Z= 8\,
v\,\Lambda_{Z}^{2}$ giving
	\be
\label{action}
S_E =  \frac{27\,\pi^2\,\tilde{S}^4}{2\,\epsilon^3} \geq
5 \times 10^{5}\,(\frac{v}{\Lambda_Z})^4 \,.
\ee
The scale $v$ of $U(1)_{``PQ''}$ breaking is unknown but even
when it is of the order of the electroweak scale, one would
still get $S_E \geq 5 \times 10^{61}$ which is huge. This will
assure us that the transition will take a very long time from
the present epoch to complete. 

{\bf (III) Dark Matter}:

There are several possibilities concerning Dark Matter candidates
in our model. To be more specific, we will concentrate on the
non-SUSY case. 

A combination
of various data (WMAP, etc..) gives a constraint for the non-baryonic
dark matter density as follows: $\Omega_M\,h^2 = 0.135_{-0.009}^{+0.008}$,
for $h \approx 0.72$ giving $\Omega_{M} \sim 0.26$.
(See \cite{silk} for an excellent recent review.)
Among the candidates
for Dark Matter, there is an interesting one that goes by the name
of WIMP (for weakly interacting massive particle). An approximate
solution to the Boltzman equation gives \cite{kamionkowski} 
$\Omega_{\chi}h^2
\sim 3 \times 10^{-27}\,cm^3\,s^{-1}/<\sigma_{A}v>$, where $\chi$
denotes a generic WIMP, $\sigma_A$ the annihilation cross section
and $v$ the relative velocity. It is generally noticed 
\cite{kamionkowski}that
$\Omega_{\chi}h^2$ is of order unity if $\sigma_{A}$ is of a typical
weak cross section, e.g. $\sigma_{A} \sim \alpha^2/m^2$ with
$\alpha \sim O(0.01)$ and $m \sim O(100\,GeV$. In brief, a stable WIMP
with a weak cross section might fulfill the requirement of being
a plausible candidate for Dark Matter.

The $\psi_{2}^{(Z)}$ particle with mass $m_2 \sim
O(100\,GeV)$ is just such a candidate. It is {\em stable} (see
the comment made in the particle physics discussion). Its
annihilation cross section is typically of the order of the weak cross
section as we have discussed above. Let us recall that the fact
that it is so is because $\alpha_{Z}(T \sim m_2) \sim \alpha_2 (m_2)$, a
notable feature of our model.

\begin{acknowledgments}
I would like to thank Paul Frampton and Marc Sher for useful discussions.
This work is supported in parts by the US Department
of Energy under grant No. DE-A505-89ER40518. 
\end{acknowledgments}

\end{document}